\documentstyle[aps,prl,epsf,twocolumn,epsfig,floats,graphicx]{revtex}

\begin{document}

\draft

\wideabs{

\title{Quasi-Langmuir-Blodgett Thin Film Deposition of Carbon Nanotubes}

\author{N.P. Armitage$^1$, J.-C.P. Gabriel$^2$ and G. Gr\"{u}ner$^{1,2}$}

\address{$^1$Department of Physics and Astronomy, University of
California, Los Angeles, CA 90095}

\address{$^2$Nanomix Inc., 5980 Horton Street, Suite 600 Emeryville, CA 94608}

\date{\today}
\maketitle
\begin{abstract}

The handling and manipulation of carbon nanotubes continues to be
a challenge to those interested in the application potential of
these promising materials.  To this end, we have developed a
method to deposit pure nanotube films over large flat areas on
substrates of arbitrary composition.  The method bears some
resemblance to the Langmuir-Blodgett deposition method used to lay
down thin organic layers.  We show that this redeposition
technique causes no major changes in the films' microstructure and
that they retain the electronic properties of as-deposited film
laid down on an alumina membrane.

\end{abstract}

}


Carbon nanotubes have emerged as materials of fundamental
importance and great application potential due to their
exceptional electrical, mechanical, and thermal properties
\cite{NanotubeR1,NanotubeR2}. Various proposals exist for their
incorporation into devices \cite{Bachtold,Martel} in single tube
or thin film architectures.  It has recently been found that
networks of nanotubes can act as conducting channels in field
effect transistors (FETs) \cite{FET1,FET2}. In addition, such
films could be used in fault tolerant chemical or biological
sensors \cite{Collins,Kong,Chen,Shim,Modi}, thermal heat shunts,
as well as in measurements of fundamental nanotube properties in
cavity and optical experiments \cite{Armitage}.  For such
applications the preparation of uniform flat films is of paramount
importance.

Although a number of methods do exist it is still far from a
trivial task to prepare thin relatively uniform films.  The
deposition of such films from solution is difficult as nanotubes
have very poor solubility in typical solvents without the use of
surfactants. Strong intertube attractions, violent hydrophobicity,
and low solubility at moderate concentrations all fight against
typical wet chemistry techniques in making uniform films.  Even if
nanotubes can be suspended at a low concentrations under certain
conditions \cite{Bahr}, simple air drying of well suspended
nanotubes on substrates results in flocculation when the local
concentration approaches the solubility limit. Surfactants that
make the nanotubes compatible with aqueous dispersions may be
inappropriate for applications that require pure nanotubes. It is
possible to deposit from dilute suspensions onto filtering
membranes, but unless other steps are taken then one is
constrained to use the filter as a substrate. This may be
inappropriate for applications where one wants to apply gate
voltages to the film (in FETs) or use optically transparent
substrates.  Under some situations, nanotubes can be deposited
with spin coating, but for thin films ($<$1 $\mu$m), it is
difficult to get adequate uniformity with such a method.  A method
has been found to create aligned thin films from evaporation of
Triton X suspensions at the air-substrate-suspension triple line,
but this technique seems to be limited to collodial dispersions
and short nanotubes \cite{Shimoda}.

Very thin nanotube films can be grown using new chemical vapor
deposition (CVD) based methods \cite{CVD1,CVD2}. However, this
latter process requires innovative catalyst deposition and high
temperatures - not compatible with CMOS technology - for the
nanotube growth. These are barriers for a nanotube-CMOS
integration process, in particular in the case of substrates that
are not heat tolerant.

In this work we describe a method of laying down thin uniform
films of carbon nanotubes on substrates of arbitrary composition
that has a number of advantages over the above described
techniques. A dilute nanotube solution is deposited onto an
alumina membrane. The volume below the membrane is then
back-filled with a fluid (typically deionized water) immiscible
with the suspending liquid; the nanotube film can then be floated
on an aqueous layer as a 'raft'. Upon drawing a substrate through
the free liquid surface a thin uniform coating of nanotubes can be
transferred.

Our method bears some resemblance to the Langmuir-Blodgett thin
film deposition used to create uniform layers of organic
molecules. In the Langmuir-Blodgett technique a hydrophobic group
enables a thin film to lay on a free surface of a subphase
compound (typically water). Although true Langmuir-Blodgett thin
films have been made with nanotubes embedded in a surfactant
matrix suspended on top of an aqueous subphase and then pulling
the substrate through the surface \cite{Krstic}, such deposition
methods are not useful if the final nanotube product need be pure.

This technique has a number of important advantages over simple
air drying of a liquid suspension of nanotubes.  As detailed
above, the liquid phase interaction between nanotubes in solution
results in large flocculation effects as a nanotube suspension
dries. Air drying gives totally unsuitable results, where
nanotubes clump themselves in concentrated 0.1-mm 'piles'.  These
kind of effects even come into play when spin-coating nanotubes
and prevent very thin films from being deposited uniformly.

\begin{figure}[htb]
 \center
  \includegraphics[width=5cm]{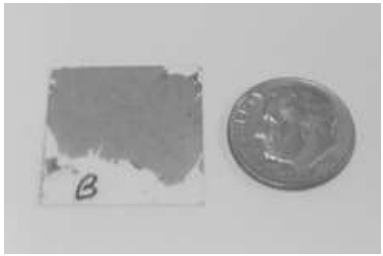}
  \vspace{0.4cm}
  \caption{Large relatively uniform films can be deposited on substrates of
  arbitrary composition.  Shown is a redeposited nanotube film on a microscope
  coverslide.}
  \label{Filmpct}
\end{figure}

High pressure carbon monoxide (HiPCO) process single wall
nanotubes were obtained from Carbon Nanotechnologies Inc and
additionally purified based on a procedure derived from Chiang et
al. \cite{Chiang}. Nanotubes, held in an alumina crucible, were
allowed to react in moist air (obtained by bubbling ambient air
through an immersed frit in room temperature water) at 225 degree
Celsius for 18h (using a tubular oven; flow = 0.1 L/min). The
remaining solid was suspended in concentrated HCl (37\%, ACS grade
Aldrich) and immersed in an ultrasound bath for 30 min. The
nanotubes were then filtered, rinsed, and washed with water until
a neutral pH was reached and then dried in a vacuum oven.

The purified nanotubes were ultrasonically dispersed for 30
minutes in 10:1 ratio of 1,2-dimethylbenzene (ortho-xylene) and
1,2-dichlorobenzene. The nanotube concentration in solution was
$\approx$ 3 mg/L.  It is a matter of some debate whether a true
\textit{solution} of nanotubes versus a \textit{suspension} can be
obtained.  As we don't make a distinction, we estimate the
saturation concentration of nanotubes in our solvent to be
approximately $\approx $15 mg/L as defined in Ref. \cite{Bahr}
which is significantly greater than our working concentration.

A specified amount of solution (for the below images - an amount
sufficient to give 8.3$\mu$g/cm$^2$ which gives an $\approx 1\mu$m
thick film) is deposited onto an alumina membrane (Whatman 0.02
$\mu$m pore size).  A vacuum can then be engaged which allows one
to remove the liquid smoothly and uniformly on a time scale short
enough that flocculation and large-scale structures do not have
time to form. We henceforth refer to the film formed on the
alumina membrane that employs it as a substrate as 'as-deposited'.
As-deposited films were also measured for comparison purpose.
After the liquid has been removed, pumping continues and the film
is allowed to dry for some additional time. Exact numbers are
difficult to give; at our vacuum pumping rate, we dried the films
for $\approx$ 1 minute after the excess liquid had been removed.
This leaves a nanotube network with no excess fluid but which is
still somewhat damp with the suspension fluid.

After deposition on the alumina membrane the chamber underneath
the film is backfilled with deionized water.  Our apparatus is
constructed such that this water can be allowed to wash up and
over the deposited nanotube film.  We have found that the wetting
of the alumina membrane and intended deposition substrate is aided
by the addition of $\approx$2\% isopropanol to the deionized
water, although under some circumstances this caused the nanotube
layer to 'shrivel'. As the fluid washes over the filter, parts of
the film, still wet with the suspending fluid, will lift off the
substrate and float on top of the water in the form of a single or
possibly multiple large 'rafts'.  The intended final deposition
substrate (glass cover slides in the below images) can be slipped
through the free surface and under the floating rafts, whereby the
vacuum can be reengaged and the water removed, redepositing the
nanotube film directly on top of the new substrate.  If the
as-deposited film did not immediately lift up, subsequent water
washes typically dislodged it.  Using the above method, we have
succeeded in laying down large scale relatively uniform films with
areas of up to 3.3cm$^2$ as seen in Fig. \ref{Filmpct}.

The nanotube suspending solution must be chosen carefully.  We
chose to use a hybrid mixture to optimize the necessary features
of high nanotube solubility, a specific gravity of less than 1,
and immiscibility in water. Two of the best solvents for nanotubes
1,2-dichlorobenzene (solubility up to 95 mg/L) and chloroform
($\approx$ 30 mg/L) \cite{Bahr} have relatively high specific
gravities (1.3 and 1.48 respectively).  Such solvents are
inappropriate as it is essential to our technique that the solvent
have a density less than that of water, otherwise the 'raft' will
not float.  A number of other candidate solvents (for instance
tetrahydrofuran and dimethylformamide) are miscible in water and
hence inappropriate. For this reason we chose to use a 10:1 ratio
of 1,2-dimethylbenzene (ortho-xylene) and 1,2-dichlorobenzene. The
solution combines the features of reasonably high nanotubes
solubility (estimated to be $\approx$ 15mg/L), low specific
gravity (0.92g/ml), and immiscibility in water.

\begin{figure} [htb]
  \center
  \includegraphics[width=6.5cm]{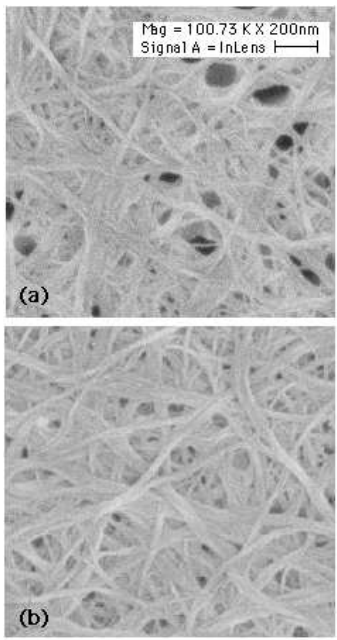}
  \vspace{0.4cm}
  \caption{SEM plots of nanotube films.  The inset shows a 200 nm reference size
  mark.  (a) Film deposited on alumina filter substrate without water immersion
  step.  (b) Film deposited with water immersion step.  The film's
  microstructure is unaffected by water immersion. }
  \label{SEM}
\end{figure}

One may have concerns regarding adverse effects the films may
suffer during their immersion and redeposition.  For instance the
extreme hydrophobicity of the nanotubes might cause them to clump
into thicker ropes upon exposure to water despite being reasonably
well dispersed prior to alumina membrane deposition. That this is
not a concern can be seen in Figs. \ref{SEM}a and \ref{SEM}b. Here
scanning electron microscope (SEM) micrographs of films, both ones
as-deposited and with the water immersion and redeposition steps,
show that the films appear to consist of well separated $\approx$6
nanotube wide 'ropes'. There is essentially no difference in their
microstructure indicating that the film morphology does not suffer
any gross effects due to water immersion.

\begin{figure} [htb]
  \center
  \includegraphics[width=6.5cm]{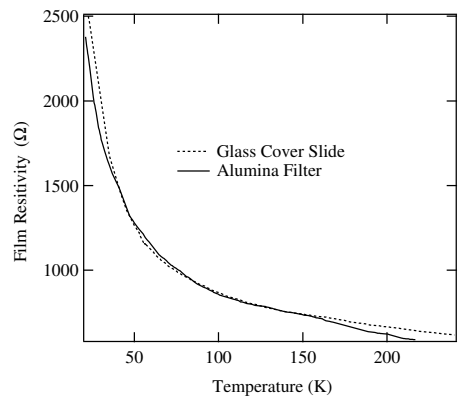}
  \vspace{0.4cm}
  \caption{DC resistivity of both as-deposited carbon nanotube films and ones
immersed in water and redeposited on glass slides.  Resistivity is
given in terms of Ohms per square of the 2D films. The small
differences between films is within the typical varibility of
as-deposited films.}
  \label{NTRes}
\end{figure}

Temperature dependent DC resistivity measurements support the
inference that there are no subtle structural changes which may
influence the film's transport properties. Data were taken on a
number of nanotube films, both as-deposited and also ones immersed
in water and then redeposited on glass slides as described above.
The measurements were done $via$ the standard 4-probe technique
with currents of 1 $\mu$A under 1 atm. of helium gas. Electrical
contact was made via Epotek silver epoxy.  In Fig. \ref{NTRes}, is
shown the resistivities of two representative films.  The
displayed data have been corrected for geometrical factors and
this is given in ohms per square unit. Both the overall magnitude
of the resistivity and its temperature dependence are unaffected
by the redeposition process. The very small differences observed
are within the typical variability of as-deposited films.

In conclusion we have described a method to deposit thin films of
carbon nanotubes on substrates of arbitrary composition. We have
shown that the structural and electronic properties of the films
are essentially unaffected by this redeposition method, opening
the way for the incorporation of such films into nanoscale
electronic devices.  While CVD growth of carbon nanotubes will
most probably remain the method of choice for nanotube-silicon
integration, the method described above may have promise in
fabricating thin films on surfaces that cannot withstand the CVD
growth environment.  As such, Quasi-Langmuir-Blodgett thin film
deposition may become an effective way for integration of networks
into transparent or plastic substrates, which would essential for
flexible electronics and display applications.

The authors would like to thank N.S. Armitage and R. Crane for
various insightful conversations and S. Kwan for help in the SWNT
purification.

\end{document}